# Quantum heat transport and effects of quantum thermal devices in noncommuting coupled spins*

CHEN Yitian, KONG Junran, LIU Huan, WANG Chen†

Department of Physics, Zhejiang Normal University, Jinhua 321004, China

**Abstract**

Quantum heat transport governs energy exchange processes and statistical laws in non-equilibrium quantum systems, and plays a pivotal role in advancing quantum thermodynamics. In this work, we comprehensively investigate the steady-state thermal transport properties of a noncommuting coupled spin system driven by a finite temperature bias. The system comprises interacting spin ensembles, each coupled to independent bosonic thermal reservoirs. We employ the quantum dressed master equation approach within the framework of open quantum system theory to accurately analyze the non-equilibrium dynamics, ensuring the validity of transport results in the strong coupling regime. Our results demonstrate that noncommuting spin coupling serves as a significant resource for modulating the nonlinearities of the heat current. Specifically, in the weak spin-coupling regime, the system exhibits robust negative differential thermal conductance (NDTC) across various spin numbers. By deriving analytical expressions for the heat current in both the single-spin and large-spin limits, we reveal that this NDTC behavior is governed by microscopic cycle fluxes. Physically, this arises because spin excitation channels induced by the cold reservoir are suppressed under a large temperature bias, thereby blocking energy exchange cycles. Conversely, in the strong spin-coupling and large temperature bias regime, the quantum system demonstrates pronounced thermal rectification. This high rectification efficiency originates from the unidirectional saturation of the heat current, rendering the system a promising candidate for high-performance thermal diodes. Furthermore, we extend the model to a three-terminal configuration to construct a quantum thermal transistor. By manipulating the temperature of the gate reservoir, we achieve efficient modulation and amplification of heat flow between the source and drain. The heat amplification factor $\beta_{\mathrm{R}}$ is shown to far exceed unity in specific operating regions, confirming significant thermal amplification. These findings not only elucidate the rich nonlinear transport phenomena induced by noncommuting interactions but also provide a theoretical foundation for designing controllable quantum thermal logic devices, such as thermal rectifiers and transistors.

---

†Corresponding author. E-mail: wangchen@zjnu.cn

‡Project supported by the Natural Science Foundation of Zhejiang Province, China (Grant No. LZ25A050001).

# 1 Introduction

Quantum heat transport is a fundamental issue in nonequilibrium physics and quantum thermodynamics[1-4]. Investigating this phenomenon is crucial for understanding the connection between microscopic quantum thermodynamic laws and quantum dynamic processes[5,6]. When the system scale enters the mesoscopic and microscopic regimes, quantum coherence, discrete energy level structures, and system-environment interactions significantly influence energy transport behavior[7,8]. Quantum heat transport is typically characterized by steady-state energy flow between thermal baths at different temperatures. Its essence is determined by the energy exchange process between the system degrees of freedom and fermionic or bosonic modes in the environment[2,4,9-13]. Since thermal environments can often be effectively described at the microscopic level as a collection of continuous-spectrum bosonic fields[14], quantum heat transport problems are usually studied within the framework of system-bosonic field coupling. This description shares high similarity with the theory of quantum light-matter interaction[15-17]. In recent years, significant progress has been made in the controllable study of nonequilibrium quantum heat transport based on experimental platforms such as superconducting circuits[18], quantum dots[19], cold atoms[20], ion traps[21,22], and solid-state spins[23].

Quantum light-matter coupled systems serve as typical models for studying quantum heat transport due to their clear structure and tunable coupling forms[8]. When multiple single-mode optical fields couple to a single qubit and are respectively coupled to bosonic thermal baths at different temperatures, the system can form a steady-state heat flow driven by the temperature difference, thereby constituting a quantum device[24-29]. Furthermore, when multiple qubits couple to a single-mode optical field to form a composite system, novel heat transport characteristics emerge[25,30,31]. Recently, research on quantum heat transport has gradually expanded to include the design and functional realization of thermal devices in typical systems such as circuit quantum electrodynamics[8], for example, thermal diodes[26,32] and three-terminal devices[33]. By introducing asymmetry in internal components of circuit quantum electrodynamics systems[34,35], inelastic scattering[36], and time-dependent and chiral modulation[37,38], one can effectively modulate heat flow asymmetry and sensitivity.

Noncommuting coupling in quantum systems mainly refers to cases where the Hermitian operators of coupling components in hybrid systems[27,39] or the operators describing system-thermal bath interactions do not commute[15,40,41]. Existing studies indicate that when a quantum system exhibits noncommuting coupling characteristics, the energy exchange process often introduces asymmetry[27,32,35,42]. This leads to features of quantum thermal devices such as thermal diodes[39], quantum thermometers[43], quantum heat suppression effects[44], and quantum thermodynamic machines[45]. One typical device model is the noncommuting coupled qubit system[39,42], where internal noncommutativity generates significant thermal rectification effects. It has been found that such noncommuting coupled quantum systems can significantly enhance the nonlinearity and asymmetry of heat transport under nonequilibrium conditions. However, other typical thermal device behaviors and their underlying physical mechanisms, such as quantum thermal transistor characteristics and microscopic circulation currents, have not yet been revealed in noncommuting coupled spin systems.

Based on the aforementioned research background, this paper investigates the quantum heat transport properties of noncommuting coupled spin systems driven by finite temperature differences using the quantum dressed-state master equation[46,47]. By analyzing the response behavior of steady-state heat flow to temperature differences and coupling parameters, we reveal the mechanism underlying the negative differential thermal conductance phenomenon. Furthermore, we demonstrate the functional characteristics of this model in terms of thermal rectification and thermal amplification. These results help deepen the understanding of nonequilibrium quantum heat transport laws in noncommuting coupled systems and provide theoretical references for the design of quantum thermal devices based on spin or qubit systems.

# 2 Model and Method

## 2.1 Noncommuting Coupled Spin Model

We consider two nonequilibrium spin ensembles coupled through a noncommuting interaction, with each ensemble independently coupled to a bosonic thermal bath, as shown in Fig. 1(a). The Hamiltonian can be expressed as

$$\hat{H}_{\mathrm{tot}} = \hat{H}_{\mathrm{S}} + \sum_{\mu=l,r} (\hat{H}_{\mathrm{B},\mu} + \hat{V}_{\mu}). \quad (1)$$

$\hat{H}_{\mathrm{S}}$ denotes the Hamiltonian of the system,

$$\hat{H}_{\mathrm{S}} = \varepsilon_{\mathrm{l}}\hat{J}_z^{\mathrm{l}} + \varepsilon_{\mathrm{r}}\hat{J}_z^{\mathrm{r}} + \lambda\hat{J}_x^{\mathrm{l}}\hat{J}_z^{\mathrm{r}}, \quad (2)$$

Here, $\varepsilon_{\mathrm{l}}$ and $\varepsilon_{\mathrm{r}}$ denote the spin splitting energies. The operator $\hat{J}_\alpha^\mu = \sum_{i=1}^{N_\mu} \hat{\sigma}_\alpha^\mu/2(\alpha = x, y, z)$ represents the collective angular momentum of the spin ensemble, where $\hat{\sigma}_\alpha^\mu$ is the Pauli operator and $N_\mu$ indicates the number of spins in the $\mu$-th ensemble. The parameter $\lambda$ signifies the noncommuting coupling strength between the two spin ensembles. The Hamiltonian of the thermal bath is given by $\hat{H}_{\mathrm{B},\mu} = \sum_k \omega_{k\mu}\hat{b}_{k\mu}^\dagger\hat{b}_{k\mu}$, where $\hat{b}_{k\mu}^\dagger$ creates a boson with frequency $\omega_{k\mu}$ in the $\mu$-th bath. The system-bath interaction is described by $\hat{V}_\mu = \hat{J}_x^\mu \sum_k (g_{k,\mu}\hat{b}_{k\mu}^\dagger + g_{k,\mu}^*\hat{b}_{k\mu})$, in which $g_{k,\mu}$ represents the coupling strength between the bath and the spins. Numerical calculations in this study employ natural units, setting the reduced Planck constant $\hbar = 1$ and the Boltzmann constant $k_{\mathrm{B}} = 1$. All physical quantities are rendered dimensionless relative to the energy level splitting $\varepsilon_{\mathrm{l}}$ of the left spin.

The coupled-system Hamiltonian $\hat{H}_{\mathrm{S}}$ admits analytical eigensolutions. Let the eigenstates be written as $\{|\phi_{\mathrm{l}}\rangle \otimes |j_{\mathrm{r}}, m_{\mathrm{r}}\rangle\}$, where the total quantum number of the right spin is $j_{\mathrm{r}} = N_{\mathrm{r}}/2$ and the magnetic quantum number is $m_{\mathrm{r}} = -N_{\mathrm{r}}/2, -N_{\mathrm{r}}/2 + 1, \cdots, N_{\mathrm{r}}/2$. The system Hamiltonian can be block-diagonalized with respect to $m_{\mathrm{r}}$. In the subspace with magnetic quantum number $m_{\mathrm{r}}$, the effective Hamiltonian is $\hat{H}_{\mathrm{S},m_{\mathrm{r}}} = (\varepsilon_{\mathrm{l}}\hat{J}_z^l + \lambda\hat{J}_x^l m_{\mathrm{r}} + \varepsilon_{\mathrm{r}} m_{\mathrm{r}})$. This yields the stationary Schrödinger equation $\hat{H}_{\mathrm{S},m_{\mathrm{r}}}|\phi_{\mathrm{l}}\rangle \otimes |j_{\mathrm{r}}, m_{\mathrm{r}}\rangle = E|\phi_{\mathrm{l}}\rangle \otimes$ $|j_{\rm r}, m_{\rm r}\rangle$. By introducing the unitary rotation operator $\hat{U} = {\rm e}^{-{\rm i}\theta \hat{J}_y^l/2}$ acting on the left subsystem and applying the unitary transformation $\hat{H}_{{\rm S},m_{\rm r}}(\theta) = \hat{U}^\dagger \hat{H}_{{\rm S},m_{\rm r}} \hat{U}$, we obtain $\hat{H}_{{\rm S},m_{\rm r}}(\theta) = \sqrt{\varepsilon_{\rm l}^2 + \lambda^2 m_{\rm r}^2} \hat{J}_z^l + \varepsilon_{\rm r} m_{\rm r}$, where the rotation angle is defined by $\tan\theta = \lambda m_{\rm r}/\varepsilon_{\rm l}$. Consequently, the eigenvalue equation for $\hat{H}_{\rm S}$ is $\hat{H}_{\rm S}|\psi_{m_{\rm l},m_{\rm r}}\rangle = E_{m_{\rm l},m_{\rm r}} \times |\psi_{m_{\rm l},m_{\rm r}}\rangle$, where the eigenvalues and eigenstates are

$$E_{m_{\rm l},m_{\rm r}} = \sqrt{\varepsilon_{\rm l}^2 + \lambda^2 m_{\rm r}^2} m_{\rm l} + \varepsilon_{\rm r} m_{\rm r}, \quad (3a)$$

$$|\psi_{m_{\rm l},m_{\rm r}}\rangle = ({\rm e}^{-{\rm i}\theta \hat{J}_y^l/2}|j_{\rm l}, m_{\rm l}\rangle) \otimes |j_{\rm r}, m_{\rm r}\rangle. \quad (3b)$$

The total quantum number of the left spin is $j_{\rm l} = N_{\rm l}/2$, and the corresponding magnetic quantum number is $m_{\rm l} = -N_{\rm l}/2, -N_{\rm l}/2+1, \cdots, N_{\rm l}/2$.

## 2.2 Quantum Master Equation

Consider the case where the system-bath coupling $\hat{V}_\mu$ is very weak. The dissipative dynamical equation for the spin system can be derived by treating $\hat{V}_\mu$ as a perturbation. Under the Born approximation, the density matrix of the total system is approximated as $\hat{\rho}_{\rm tot}(t) \approx \hat{\rho}_{\rm s}(t) \otimes \hat{\rho}_{\rm B}$, where $\hat{\rho}_{\rm s}(t)$ denotes the system density matrix. The bath remains in a local thermal equilibrium state $\hat{\rho}_{\rm B} = \prod_{\mu={\rm L,R}} \quad ({\rm e}^{-\beta_\mu H_{{\rm B},\mu}}/Z_\mu)$, with the partition function defined as $Z_\mu = {\rm Tr}_{\rm B}\{{\rm e}^{-\beta_\mu H_{{\rm B},\mu}}\}$. Here, $\beta_\mu = 1/(k_{\rm B}T_\mu)$ represents the inverse temperature, and $k_{\rm B}$ is the Boltzmann constant. Furthermore, by applying the Markov approximation, we obtain the quantum dressed-state master equation[46-48] for long-time evolution:

$$\frac{\rm d}{{\rm d}t}\hat{\rho}_{\rm S}(t) = {\rm i}[\hat{\rho}_{\rm S}(t), \hat{H}_{\rm S}] + \sum_{n,m,\mu}\{\Gamma_{nm,\mu}^{+}\hat{\mathcal{L}}_{nm}[\hat{\rho}_{\rm S}(t)] + \Gamma_{nm,\mu}^{-}\hat{\mathcal{L}}_{nm}[\hat{\rho}_{\rm S}(t)]\}, \quad (4)$$

Here, the dissipation operator is expressed as

$$\hat{\mathcal{L}}_{nm}[\hat{\rho}_{\rm S}(t)] = |\psi_n\rangle\langle\psi_m|\hat{\rho}_{\rm S}(t)|\psi_m\rangle\langle\psi_n| - (|\psi_m\rangle\langle\psi_m|\hat{\rho}_{\rm S}(t) + \hat{\rho}_{\rm S}(t)|\psi_m\rangle\langle\psi_m|)/2;$$

$\Gamma^{\pm}_{nm,\mu}$ denotes the incoherent transition rate,

$$\Gamma^{+}_{nm,\mu} = \gamma_{\mu}(\Delta_{nm}) n_{\mu}(\Delta_{nm}) |\langle\psi_n|\hat{J}^{\mu}_{x}|\psi_m\rangle|^2, \text{(5a)}$$

$$\Gamma^{-}_{nm,\mu} = \gamma_{\mu}(\Delta_{nm})[1 + n_{\mu}(\Delta_{nm})] |\langle\psi_n|\hat{J}^{\mu}_{x}|\psi_m\rangle|^2. \text{(5b)}$$

$\Gamma^{\pm}_{nm,\mu}$ describes the process in which the $\mu$th thermal bath participates in the transition of the system from state $|\psi_m\rangle$ to $|\psi_n\rangle$ by absorbing or emitting a boson with energy $\Delta_{nm}$. Here, the energy level spacing is defined as $\Delta_{nm} = E_n - E_m$, where $E_n$ represents the eigenenergy from Eq. (3a). The term $\gamma_{\mu}(\Delta_{nm})$ denotes the spectral function characterizing the coupling strength between the system and the environment, while $|\langle\psi_n|\hat{J}^{\mu}_{x}|\psi_m\rangle|^2$ represents the transition coefficient. Additionally, $n_{\mu}(\Delta_{nm}) = 1/(e^{\beta_{\mu}\Delta_{nm}} - 1)$ indicates the Bose-Einstein distribution function. Let $P_n$ be the population of the $n$th eigenlevel when the system reaches a steady state. We define the steady-state heat flow $J_{\mu}$ into the $\mu$th thermal bath as the energy exchange rate between the system and that bath per unit time. In Appendix A, we derive the expression for the steady-state heat flow into the $\mu$th thermal bath by analyzing the microscopic energy interaction processes via the dressed-state master equation.

$$J_{\mu} = \sum_{n>m} \Delta_{nm} (\Gamma^{-}_{nm,\mu} P_n - \Gamma^{+}_{nm,\mu} P_m). \qquad (6)$$

Under steady-state conditions, the heat flow $J_{\mathrm{l}}$ into the left thermal bath and the heat flow $J_{\mathrm{r}}$ into the right thermal bath satisfy $J_{\mathrm{l}} = -J_{\mathrm{r}}$ according to the law of conservation of energy. Therefore, the magnitude of $J_{\mathrm{r}}$ characterizes the overall heat transport capacity of the system between the two thermal baths. The following discussion focuses primarily on $J_{\mathrm{r}}$.

# 3 Steady-State Transport

This section analyzes the behavior of steady-state heat flow for different spin numbers as functions of the thermal bath temperature difference and spin coupling strength. We observe non-monotonic changes in energy flow under weak coupling. Furthermore, we investigate the thermal rectification and thermal amplification effects in noncommuting coupled spin models.

## 3.1 Steady-State Heat Flow

Figure 1(b) shows that under weak spin-spin coupling strength ($\lambda = 0.01$), the steady-state heat flow $J_{\mathrm{r}}$ initially increases and then decreases as the temperature difference $\delta T$ increases. This indicates that heat flow is suppressed in the large temperature difference regime, corresponding to the emergence of negative differential thermal conductance[49,50]. This behavior is clearly visible for different numbers of spins where $N_\mathrm{l} = N_\mathrm{r}$. This demonstrates that negative differential thermal conductance is not limited to small-scale systems but is a universal feature of steady-state heat transport in noncommuting coupled spin systems. Additionally, the peak heat flow increases monotonically with the number of spins. Therefore, increasing the number of spins helps enhance the heat flow signal.

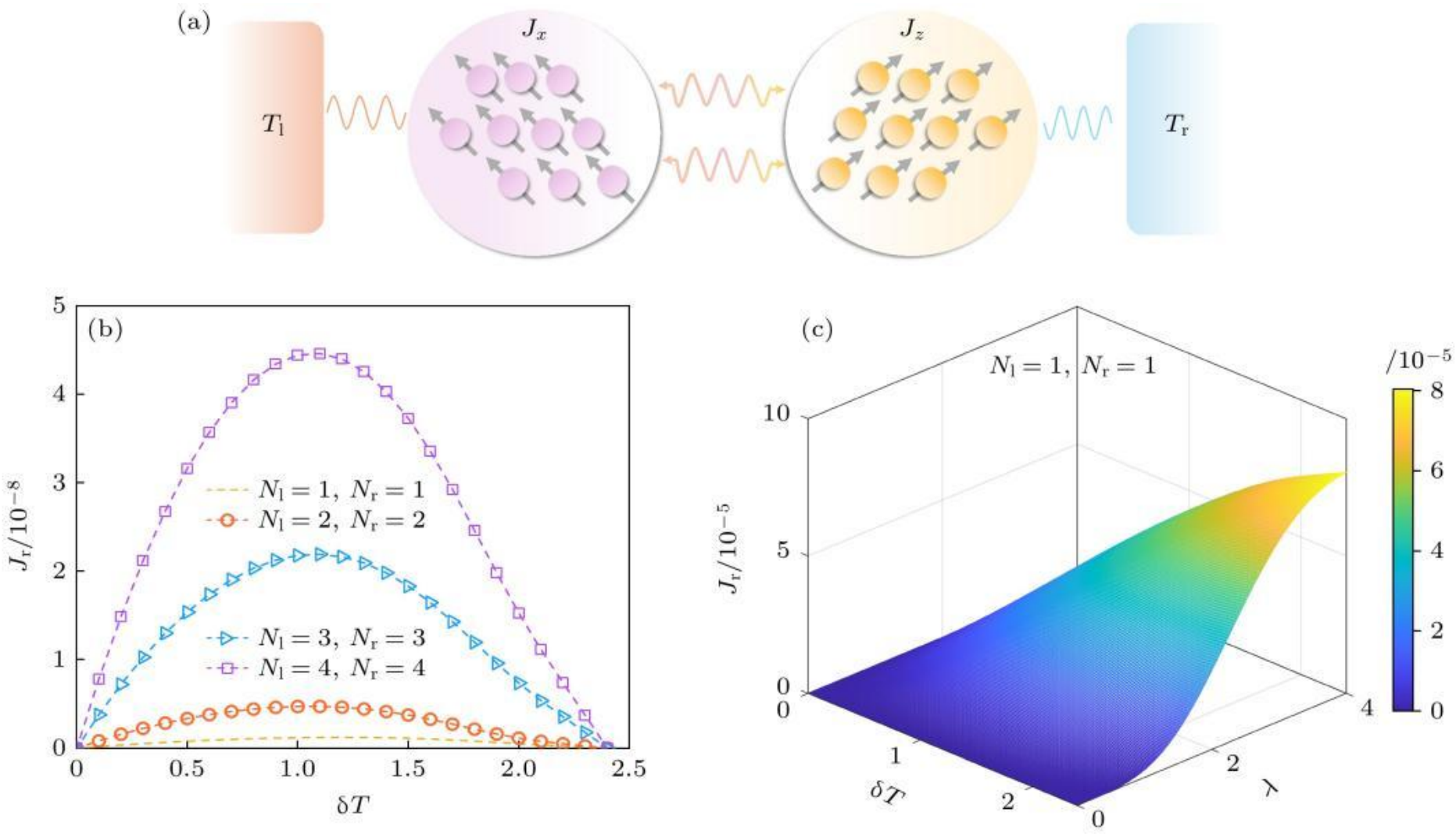

**Fig. 1** (a) **Schematic illustration of the noncommuting coupled spin model under finite thermodynamic bias; (b) steady-state energy current $J_\mathrm{r}$ as a function of temperature bias $\delta T$ for various spin numbers ($N_\mathrm{l}, N_\mathrm{r}$), with the spin-spin coupling strength set to $\lambda = 0.01$; (c) effects of $\delta T$ and $\lambda$ on $J_\mathrm{r}$, where $N_\mathrm{l} = N_\mathrm{r} = 1$. Other system parameters are $\varepsilon_\mathrm{l} = \varepsilon_\mathrm{r} = 1$, $\alpha_\mathrm{l} = \alpha_\mathrm{r} = 0.001$, $\omega_\mathrm{c} = 10$, $T_\mathrm{l} = T_0 + \delta T/2$, $T_\mathrm{r} = T_0 - \delta T/2$, and $T_0 = 1.2$. In this study, $\lambda$ and $\omega_\mathrm{c}$ are scaled by $\varepsilon_\mathrm{l}$. The steady-state heat current $J_\mathrm{r}$ is dimensionless, in units of $\hbar\varepsilon_\mathrm{l}^2$. Temperatures $T_{\mathrm{l(r)}}$ and temperature bias $\delta T$ are scaled by $\hbar\varepsilon_\mathrm{l}/k_\mathrm{B}$.**

We further analyze the behavior of heat flow from an analytical perspective. Appendix B discusses small spin systems with $N_\mathrm{l} = N_\mathrm{r} = 1$. This analysis yields an analytical expression for the steady-state heat flow under weak spin-spin coupling:

$$J_\mathrm{r} = \left(\frac{\lambda}{4\varepsilon_\mathrm{l}}\right)^2 \varepsilon_\mathrm{r} \sum_{q=\pm 1} \frac{\gamma_\mathrm{r}(\varepsilon_\mathrm{r}+q\varepsilon_\mathrm{l})}{1-2f_\mathrm{r}(\varepsilon_\mathrm{r}+q\varepsilon_\mathrm{l})} \times \{[1-f_\mathrm{r}(\varepsilon_\mathrm{r}+q\varepsilon_\mathrm{l})]f_\mathrm{l}(q\varepsilon_\mathrm{l})f_\mathrm{r}(\varepsilon_\mathrm{r}) - f_\mathrm{r}(\varepsilon_\mathrm{r}+q\varepsilon_\mathrm{l})[1-f_\mathrm{l}(q\varepsilon_\mathrm{l})][1+f_\mathrm{r}(\varepsilon_\mathrm{r})]\}, \quad (7)$$

Here, $q$ is the summation index, taking values of $\pm 1$; the distribution functions are $f_\mu(\omega) = 1/[1+\exp(\beta_\mu\omega)]$ and $f_\mu(-\omega) = 1 - f_\mu(\omega)$; $\varepsilon_{\mathrm{l(r)}}$ denotes the splitting energy of spin $l(r)$; and $\gamma_\mathrm{r}(\omega)$ represents the spectral function for the coupling between the system and the right thermal bath. Equation (7) reveals that $J_\mathrm{r}$ consists of two cyclic currents with opposite directions. As $\delta T \to 2T_0$, the temperature of the right bath approaches zero (i.e., $T_{\mathrm{r}} \to 0$), causing the right bath distribution function to satisfy $f_{\mathrm{r}}(\omega) \to 0$. Consequently, the upward excitation processes in the spin system involving $f_{\mathrm{r}}(\varepsilon_{\mathrm{r}})$ and $f_{\mathrm{r}}(\varepsilon_{\mathrm{r}} \pm \varepsilon_{\mathrm{l}})$ in Eq. (7) are significantly suppressed. Although the left bath remains at a high temperature, its thermal excitation cannot sustain continuous energy transport. Due to the lack of effective energy exchange channels between the system and the right bath, the steady-state heat flow inevitably decays and ultimately approaches zero in the strong non-equilibrium limit where $\delta T \to 2T_0$. In the regime with a large number of spins, analytical solutions for weak spin coupling can also be derived via perturbation analysis, as detailed in Appendix C:

$$J_{\mathrm{r}} = \left(\frac{\lambda\sqrt{N_{\mathrm{l}}N_{\mathrm{r}}}}{4\varepsilon_{\mathrm{l}}}\right)^2 \varepsilon_{\mathrm{r}} \sum_{q=\pm1} \gamma_{\mathrm{r}}(\varepsilon_{\mathrm{r}} + q\varepsilon_{\mathrm{l}}) \times \{[1 + n_{\mathrm{r}}(\varepsilon_{\mathrm{r}} + q\varepsilon_{\mathrm{l}})] n_{\mathrm{l}}(q\varepsilon_{\mathrm{l}}) n_{\mathrm{r}}(\varepsilon_{\mathrm{r}}) - n_{\mathrm{r}}(\varepsilon_{\mathrm{r}} + q\varepsilon_{\mathrm{l}})[1 + n_{\mathrm{l}}(q\varepsilon_{\mathrm{l}})][1 + n_{\mathrm{r}}(\varepsilon_{\mathrm{r}})]\}, \quad (8)$$

Here, the distribution function is given by $n_\mu(\omega) = 1/[\exp(\beta_\mu \omega) - 1]$, with $n_\mu(-\omega) = [1 + n_\mu(\omega)]$. This indicates that multi-spin systems can enhance collective energy transport capacity ($J_{\mathrm{r}} \propto N_{\mathrm{l}} N_{\mathrm{r}}$). However, as $\delta T \to 2T_0$, the temperature of the right heat bath approaches zero, causing its distribution function to satisfy $n_{\mathrm{r}}(\omega) \to 0$. Consequently, both $n_{\mathrm{r}}(\varepsilon_{\mathrm{r}})$ and $n_{\mathrm{r}}(\varepsilon_{\mathrm{r}} \pm \varepsilon_{\mathrm{l}})$ in Eq. (8) tend toward zero, leading to a natural decay of the heat flow to zero. Under these conditions, the incoherent transition channels induced by noncommuting coupling remain dominated by cyclic currents. This ultimately limits the continuous growth of steady-state heat flow, ensuring that negative differential thermal conductance behavior remains robust even with large spin numbers.

We further investigated heat flow behavior under strong spin coupling. As spin coupling increases, Figure 1(c) shows that heat flow monotonically enhances with $\lambda$ under finite temperature difference conditions. This behavior differs significantly from that observed under weak spin coupling. It suggests that strong noncommuting coupling effectively improves energy exchange capacity between the system and the heat baths.

So far, we have mainly considered the case in which the two spin ensembles have equal numbers ($N_{\mathrm{l}} = N_{\mathrm{r}}$). Correspondingly, negative differential thermal conductance appears in the steady-state transport of noncommuting coupled spin systems. We now discuss scenarios where $N_{\mathrm{l}}$ and $N_{\mathrm{r}}$ are unequal. Figure 2 demonstrates that negative differential thermal conductance persists even when the spin numbers in the left and right quantum subsystems differ. Furthermore, the temperature difference corresponding to the optimal heat flow shifts: (a) when $N_{\mathrm{r}} > N_{\mathrm{l}}$, the optimal temperature difference decreases; (b) when $N_{\rm l} > N_{\rm r}$, the optimal temperature difference increases. Overall, this analysis indicates that negative differential thermal conductance in noncommuting coupled spin systems is a universal behavior.

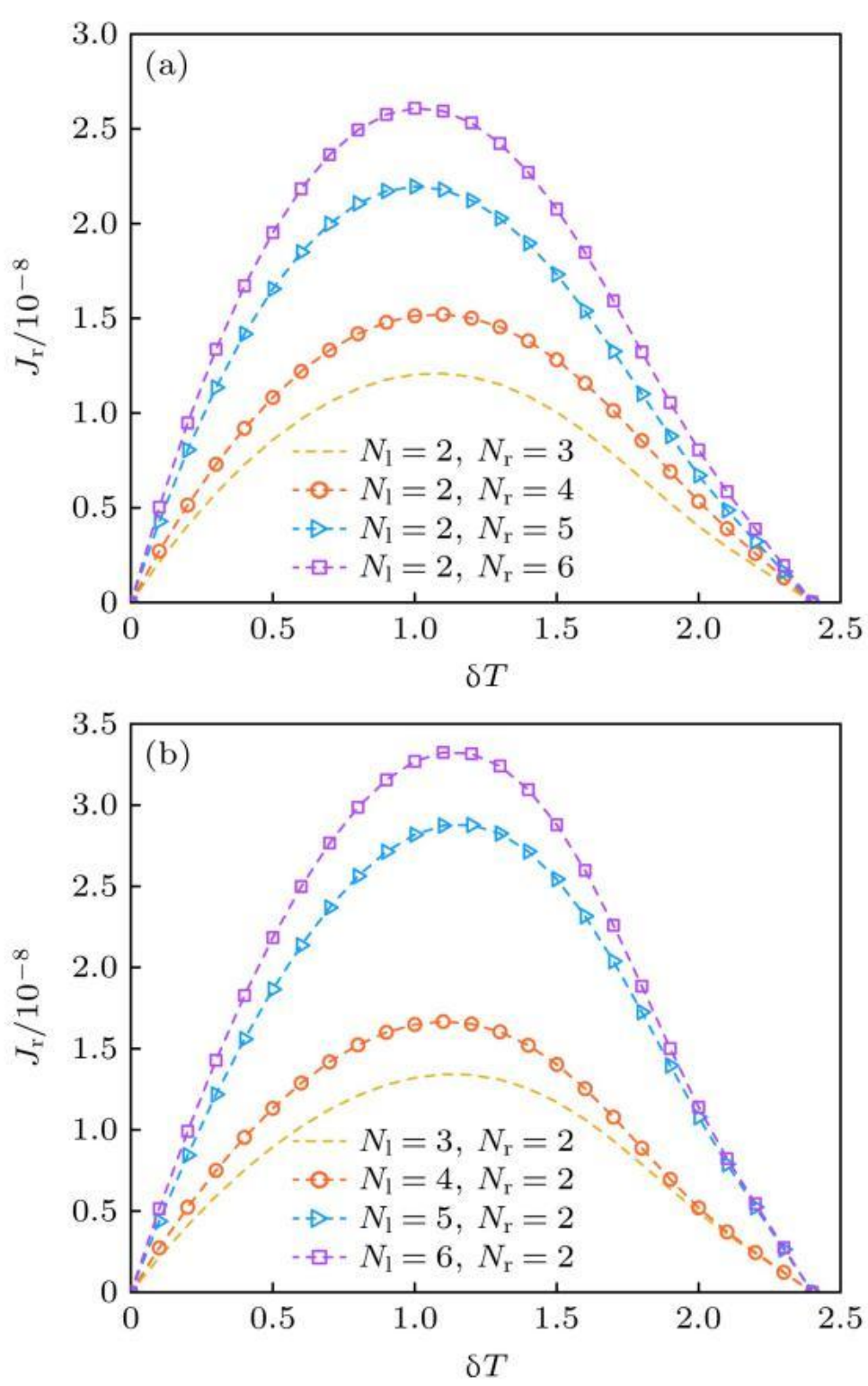

**Fig. 2 Nonmonotonic behavior of the steady-state heat current $J_{\rm r}$ with increasing temperature difference: (a) fixing $N_{\rm l}=2$ while continuously increasing $N_{\rm r}$; (b) fixing $N_{\rm r}=2$ while continuously increasing $N_{\rm l}$. Other system parameters are set as $\varepsilon_{\rm l}=\varepsilon_{\rm r}=1$, $\lambda=0.01$, $\alpha_{\rm l}=\alpha_{\rm r}=0.001$, $\omega_{\rm c}=10$, $T_{\rm l}=T_0+\delta T/2$, $T_{\rm r}=T_0-\delta T/2$, and $T_0=1.2$. In this study, $\lambda$ and $\omega_{\rm c}$ are expressed in units of $\varepsilon_{\rm l}$. The steady-state heat current $J_{\rm r}$ is a dimensionless quantity in units of $\hbar\varepsilon_{\rm l}^2$, while the temperatures $T_{\rm l(r)}$ and temperature difference $\delta T$ are expressed in units of $\hbar\varepsilon_{\rm l}/k_{\rm B}$.**

## 3.2 Thermal Device Effects in Quantum Systems

Thermal device effects in quantum systems have recently attracted significant attention from researchers[34,51-54]. Typical device effects include thermal rectification and thermal amplification. Under finite temperature differences, the thermal rectification effect in two-terminal systems can be characterized by the thermal rectification factor[50,55]

$$R=\frac{|J_{+}+J_{-}|}{max(|J_{+}|,|J_{-}|)}, \qquad (9)$$

Here, $J_{+}$ denotes the steady-state heat current under a temperature difference between the two terminals, whereas $J_{-}$ represents the steady-state current when the temperatures of the two terminals are swapped. As the system approaches symmetric heat transport ($J_+ \approx -J_-$), the rectification factor $R$ approaches zero. Conversely, when $R \to 1$, the system exhibits significant thermal rectification characteristics.

Figure 3 illustrates the variation of the rectification factor $R$ with the temperature difference $\delta T$ and the coupling strength $\lambda$ for different numbers of spins, where $N_\mathrm{l} = N_\mathrm{r}$. In the small temperature difference regime, the rectification factor remains generally low, indicating that heat transport in the spin system is approximately symmetric. As the temperature difference $\delta T$ increases, $R$ shows a marked enhancement. This suggests that a larger temperature bias helps amplify the difference between forward and reverse heat currents. This effect arises from the negative differential thermal conductance behavior under positive temperature differences, as shown in Figure 1(b). Particularly in the large temperature difference limit, $J_+ \approx 0$, while the reverse heat current remains finite, directly leading to $R \approx 1$. On the other hand, the rectification factor exhibits a clear non-monotonic dependence on the coupling strength $\lambda$. For a small number of spins (e.g., $N_\mathrm{l} = N_\mathrm{r} = 1$), as shown in Figure 3(a), $R$ is small under strong coupling, indicating that the rectification effect is somewhat suppressed. When the number of spins increases (e.g., $N_\mathrm{l} = N_\mathrm{r} = 4$), as shown in Figures 3(b)—(d), the rectification factor increases significantly and can approach 1 in certain regions. Figure 4 shows that negative differential thermal conductance is absent in this case; however, the heat current in the positive temperature difference region exhibits saturation characteristics as the temperature difference increases. This indicates that multi-spin coupling effects can effectively amplify the asymmetry of heat transport under non-equilibrium conditions.

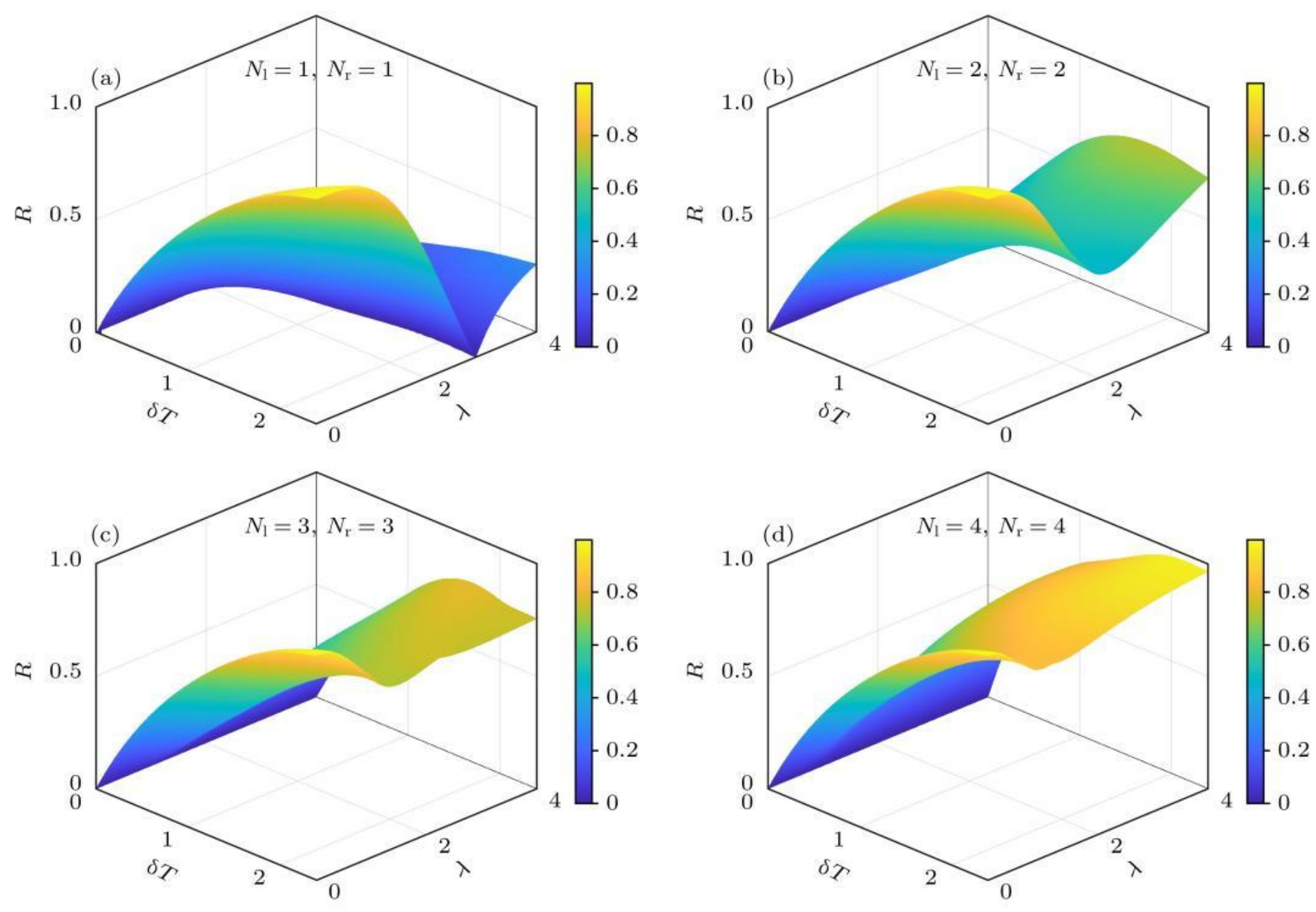

**Fig. 3 Behavior of the rectification factor $R$ as a function of temperature bias and spin coupling strength for different numbers of spins: (a) $N_{\mathrm{l}} = N_{\mathrm{r}} = 1$; (b) $N_{\mathrm{l}} = N_{\mathrm{r}} = 2$; (c) $N_{\mathrm{l}} = N_{\mathrm{r}} = 3$; (d) $N_{\mathrm{l}} = N_{\mathrm{r}} = 4$. Other system parameters are set as $\varepsilon_{\mathrm{l}} = \varepsilon_{\mathrm{r}} = 1$, $\alpha_{\mathrm{l}} = \alpha_{\mathrm{r}} = 0.001$, $\omega_{\mathrm{c}} = 10$, $T_{\mathrm{l}} = T_0 + \delta T/2$, $T_{\mathrm{r}} = T_0 - \delta T/2$, and $T_0 = 1.2$. Here, $\lambda$ and $\omega_{\mathrm{c}}$ are in units of $\varepsilon_{\mathrm{l}}$, while the temperatures $T_{\mathrm{l(r)}}$ and temperature bias $\delta T$ are in units of $\hbar\varepsilon_{\mathrm{l}}/k_{\mathrm{B}}$.**

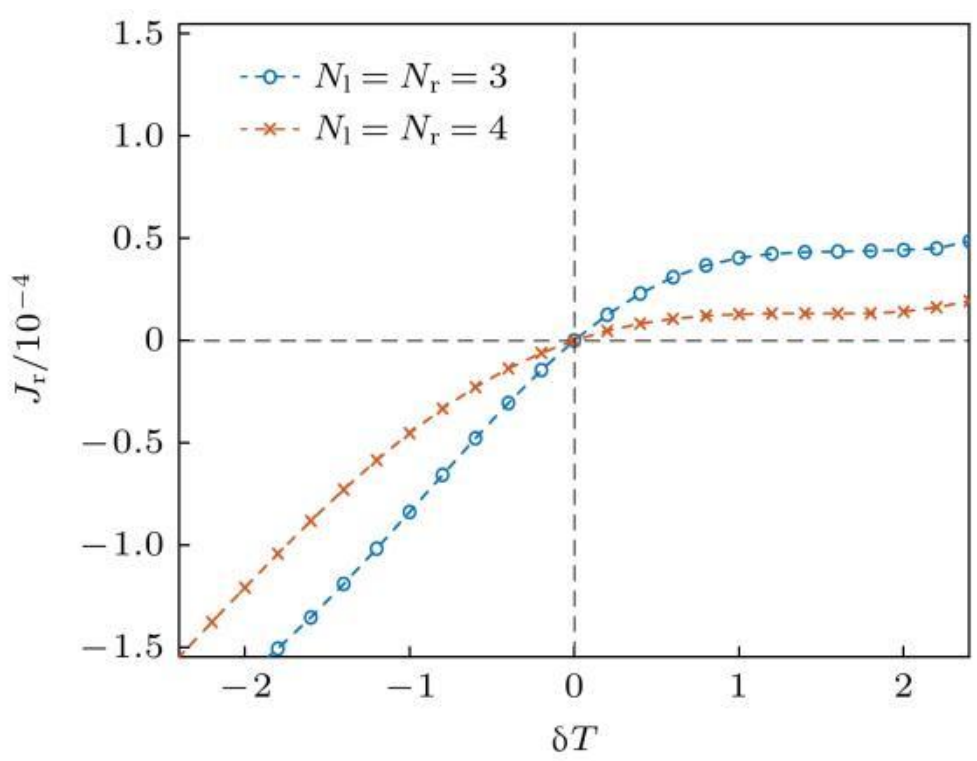

**Fig. 4 Steady-state heat current $J_{\mathrm{r}}$ versus temperature difference $\delta T$ at strong spin-spin coupling ($\lambda = 0.2$). The blue dashed line with circles corresponds to $N_{\mathrm{l}} = N_{\mathrm{r}} = 3$, and the brown dashed line with crosses corresponds to $N_{\mathrm{l}} = N_{\mathrm{r}} = 4$. Other system parameters are given by $\varepsilon_{\mathrm{l}} = \varepsilon_{\mathrm{r}} = 1$, $\alpha_{\mathrm{l}} = \alpha_{\mathrm{r}} = 0.001$, $\omega_{\mathrm{c}} = 10$, $T_{\mathrm{l}} = T_0 + \delta T/2$, $T_{\mathrm{r}} = T_0 - \delta T/2$, and $T_0 = 1.2$. In particular, $\lambda$ and $\omega_{\mathrm{c}}$ are scaled by $\varepsilon_{\mathrm{l}}$. The steady-state heat current $J_{\mathrm{r}}$ is dimensionless, in units of $\hbar\varepsilon_{\mathrm{l}}^2$. Temperatures $T_{\mathrm{l(r)}}$ and temperature bias $\delta T$ are scaled by $\hbar\varepsilon_{\mathrm{l}}/k_{\mathrm{B}}$.**

Thermal amplification constitutes the core effect for constructing functional quantum thermal transistors[51]. Its physical essence lies in using slight heat flow variations in one branch to control the output of stronger heat current variations in another. Based on this concept, we construct a three-port model of a noncommuting coupled spin system, as shown in Figure 5. The Hamiltonian of this quantum system is

$$\hat{H}_{\mathrm{S}} = \varepsilon_{\mathrm{l}} \hat{J}_{z}^{\mathrm{l}} + \sum_{n=1,2} \varepsilon_{\mathrm{r}}^{n} \hat{J}_{z,n}^{\mathrm{r}} + \sum_{n=1,2} \lambda_{n} \hat{J}_{x}^{\mathrm{l}} \hat{J}_{z,n}^{\mathrm{r}}. \quad (10)$$

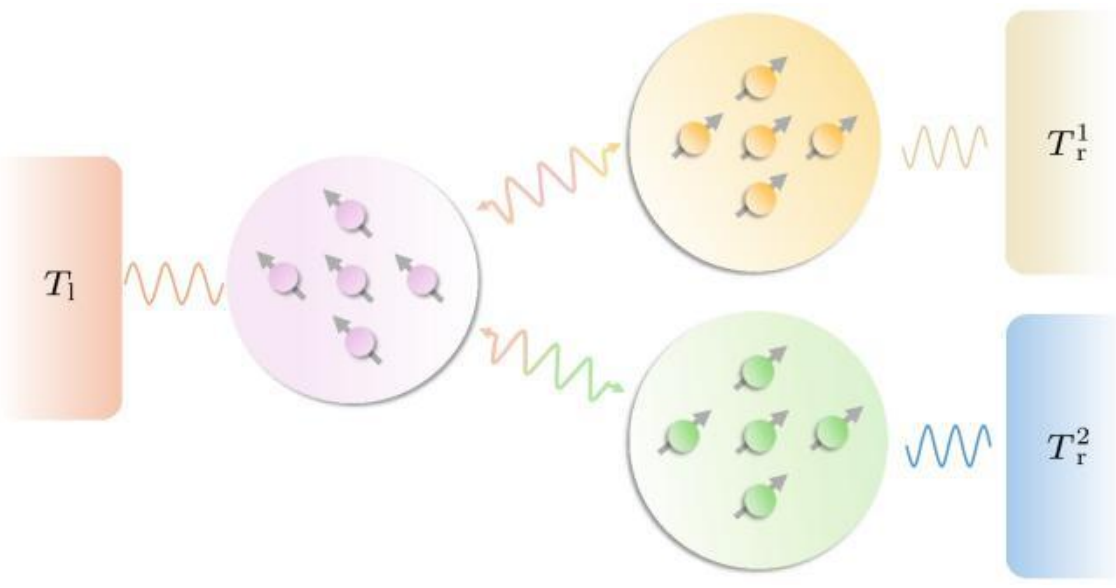

**Fig. 5 Schematic representation of a quantum thermal transistor based on a noncommuting coupled spin system: $T_{\mathrm{l}}$ denotes the temperature of the left source reservoir, whereas $T_{\mathrm{r}}^1$ and $T_{\mathrm{r}}^2$ correspond to the temperatures of the gate and drain reservoirs, respectively. Modulating $T_{\mathrm{r}}^1$ while maintaining constant $T_{\mathrm{l}}$ and $T_{\mathrm{r}}^2$ regulates the heat currents, thereby inducing an amplification effect.**

The system consists of a left-side heat source and two right-side heat terminals. The left heat bath serves as the source terminal, maintaining a constant temperature of $T_{\mathrm{l}}$. On the right side, $T_{\mathrm{r}}^{1}$ acts as the gate terminal temperature, while $T_{\mathrm{r}}^{2}$ serves as the drain terminal temperature. By fixing $T_{\mathrm{l}}$ and $T_{\mathrm{r}}^{2}$ and adjusting $T_{\mathrm{r}}^{1}$, one can effectively modify the internal heat exchange process. This approach enables the regulation of heat current at the source and drain terminals. Such a configuration provides the necessary structural basis for the thermal amplification effect. To quantitatively characterize the amplification performance of this thermal device, we define the thermal amplification factor[50]

$$\beta_{\mathrm{R}} = \left| \frac{\partial J_{\mathrm{r}}^{2}}{\partial J_{\mathrm{r}}^{1}} \right|. \quad (11)$$

When $\beta_{\mathrm{R}} > 1$, a small variation in the gate heat current produces a larger variation in the drain heat current, indicating thermal amplification. When $\beta_{\mathrm{R}} \gg 1$, a significant thermal amplification effect is realized.

Figure 6 illustrates the variation of the amplification factor $\beta_{\mathrm{R}}$ with the gate terminal temperature $T_{\mathrm{r}}^{1}$. Numerical results indicate that $\beta_{\mathrm{R}}$ exhibits pronounced nonlinear enhancement within specific parameter ranges. Notably, distinct peaks in the thermal amplification factor emerge near certain temperatures. The presence of these peaks clearly confirms the existence of the thermal amplification effect in this quantum system. A comparison between Figure 6(a) and Figure 6(b) reveals that the thermal amplification behavior persists as the number of spins in the subsystem increases. Furthermore, the amplification factor becomes more sensitive to temperature variations. Additionally, adjusting the number of spins and the coupling strength $\lambda$ causes a significant shift in the effective operating temperature range of the thermal amplification effect. This demonstrates that the thermal amplification characteristics can be flexibly tuned by modifying the internal coupling strength and system size of the quantum system. Such pronounced thermal amplification provides theoretical support for developing sensitive quantum thermal devices.

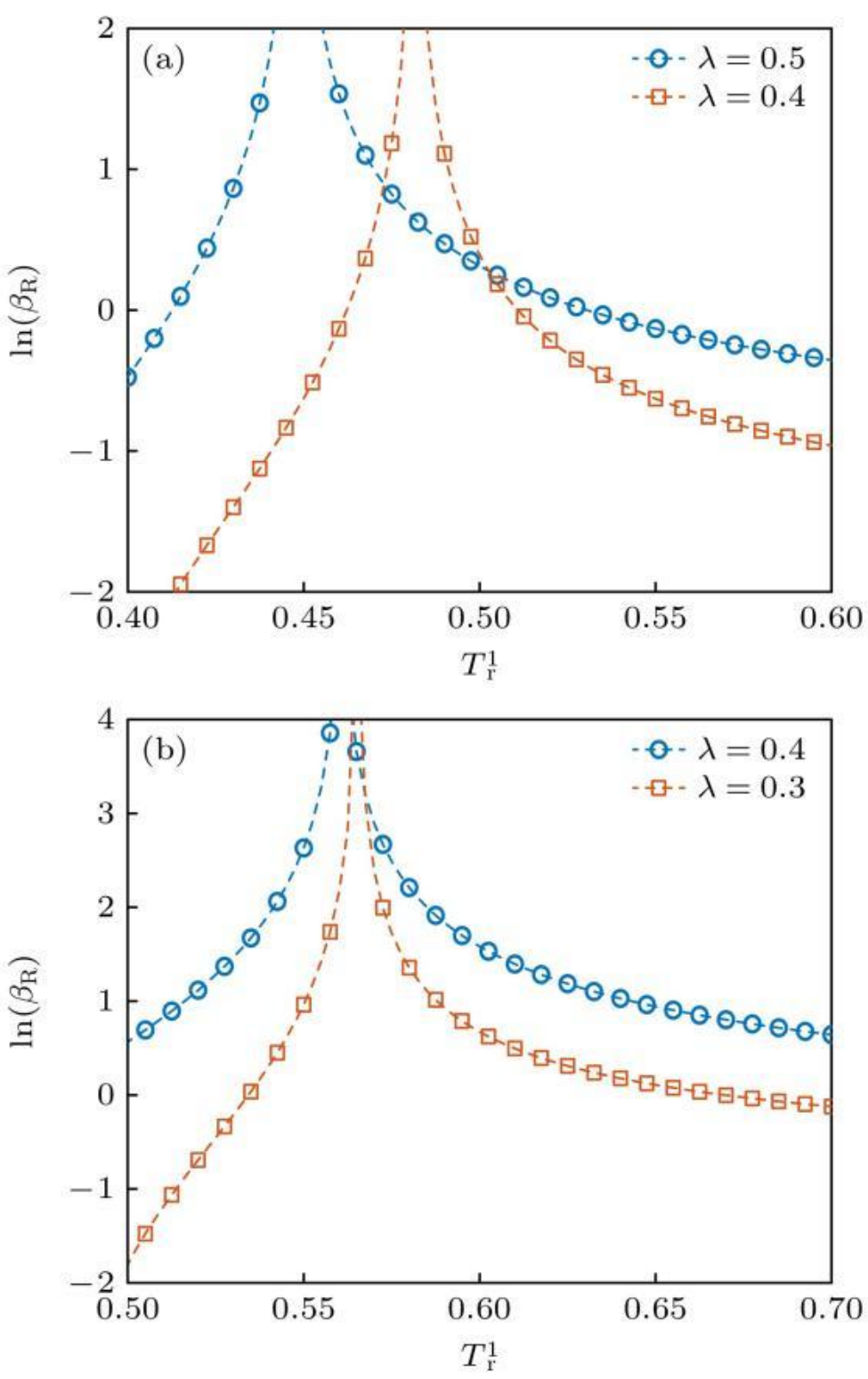

**Fig. 6 Thermal amplification factor $\beta_R$ as a function of the gate-reservoir temperature $T_r^1$: (a) $N_l = N_r^1 = N_r^2 = 1$, $\lambda_1 = \lambda/4$, $\lambda_2 = \lambda$; (b) $N_l = N_r^1 = N_r^2 = 2$, $\lambda_1 = \lambda/4$, $\lambda_2 = \lambda$. Other system parameters are given by $\varepsilon_l = \varepsilon_r^1 = \varepsilon_r^2 = 1$, $\alpha_l = \alpha_r^1 = \alpha_r^2 = 0.001$, $\omega_c = 10$, $T_l = 1.2$, and $T_r^2 = 0.2$. In particular, $\varepsilon_r^{1(2)}$, $\lambda$, and $\omega_c$ are scaled by $\varepsilon_l$. Temperatures $T_l$ and $T_r^{1(2)}$ are scaled by $\hbar\varepsilon_l/k_B$.**

# 4 Conclusion

This study systematically investigates the quantum heat transport properties and quantum thermal device effects of noncommuting coupled spin systems using the quantum dressed-state master equation method. The results reveal that negative differential thermal conductance is universally present in these systems under weak spin coupling. Furthermore, this negative differential thermal conductance behavior demonstrates robustness across various spin system sizes. Analytical expressions for heat flow are derived in the limits of single-spin and large-spin systems. The heat flow is dominated by microscopic cyclic currents. The underlying microscopic physical mechanism is as follows: in the limit of large temperature differences, spin excitation transitions induced by the low-temperature thermal bath are significantly suppressed. This suppression interrupts the cyclic energy exchange within the coupled spin system. Collective effects in multi-spin systems cause the heat flow intensity to be proportional to the product of the number of left and right spins ($J_r \propto N_l N_r$). Additionally, the noncommuting weak coupling achieves efficient thermal rectification due to unidirectional negative differential thermal conductance. In the strong coupling regime, the thermal rectification factor $R$ increases significantly with both temperature difference and spin number. Compared to the single-spin case, multi-spin coupling effects greatly enhance the asymmetry of heat transport, causing the rectification factor to approach 1 over a wide temperature range. This is primarily attributed to the unidirectional saturation of heat flow with respect to temperature difference. These findings demonstrate the potential of this system as a high-performance spin-based thermal diode. Finally, the quantum thermal transistor constructed based on a three-port noncommuting coupled spin model exhibits significant thermal amplification behavior. The study confirms that the thermal amplification factor $\beta_R$ is much greater than 1 in specific operating ranges. This enables efficient regulation of drain heat flow using weak fluctuations in gate heat flow. By optimizing the internal coupling strength $\lambda$ and the spin system size, the effective operating range of thermal amplification characteristics can be flexibly controlled. It is also worth noting that the noncommuting coupled spin model constructed in this work may hold significant potential value in the broader field of quantum thermodynamics. First, given its enhanced energy exchange capability in the strong coupling regime, this model shows promise as a novel working substance for constructing quantum heat engines and refrigerators[7,56,57]. Second, considering the significant nonlinear and fluctuation characteristics exhibited by the system far from equilibrium, this model provides an ideal physical platform for verifying thermodynamic uncertainty relations and their modified forms under noncommuting and strong coupling conditions[58-60]. This study not only reveals rich nonlinear quantum heat transport phenomena but also provides important theoretical support for developing size-tunable quantum thermal logic elements.

# Appendix A Derivation of Heat Flow Formula

According to the dressed-state master equation in Eq. (4), the total heat flow out of all thermal baths $(-\sum_{\mu=l,r} J_\mu)$ equals the expectation value of the energy change rate of the system Hamiltonian $\hat{H}_S$ caused by dissipation channels through these baths. Based on the first law of thermodynamics, this can be expressed as $\sum_\mu J_\mu = -\mathrm{Tr}_S(\hat{H}_S \dot{\hat{\rho}}_S(t))$. Therefore, the heat flow into the $\mu$-th thermal bath is defined as

$$J_\mu = -\mathrm{Tr}_S(\hat{H}_S \mathcal{D}_\mu[\hat{\rho}_S]), \quad \text{(A1)}$$

Here, $\mathcal{D}_\mu[\hat{\rho}_S]$ denotes the dissipative contribution induced solely by the $\mu$-th thermal bath in the master equation (4). We perform the analysis in the system eigenbasis $\{|\psi_n\rangle\}$. By employing the eigenvalue equation $\hat{H}_S|\psi_n\rangle = E_n|\psi_n\rangle$, we expand the trace operation in the eigenbasis as follows

$$J_\mu = -\sum_k \langle\psi_k|\hat{H}_S\mathcal{D}_\mu[\hat{\rho}_S]|\psi_k\rangle = -\sum_k E_k\langle\psi_k|\mathcal{D}_\mu[\hat{\rho}_S]|\psi_k\rangle.\text{(A2)}$$

Based on the definition of the Liouvillian dissipative operator $\hat{\mathcal{L}}_{nm}[\hat{\rho}_S]$ in Eq. (4) and the definitions of the incoherent transition rates $\Gamma^{\pm}_{nm,\mu}$ in Eqs. (5a) and (5b), summing over all energy levels yields

$$\begin{aligned} J_\mu = & -\sum_{n>m}[\Gamma^{+}_{nm,\mu}(E_nP_m - E_mP_m) \\ & +\Gamma^{-}_{nm,\mu}(E_nP_n - E_mP_n)] \qquad \text{(A3)} \\ = & \sum_{n>m}(E_n - E_m)(\Gamma^{-}_{nm,\mu}P_n - \Gamma^{+}_{nm,\mu}P_m). \end{aligned}$$

Let the energy level spacing be $\Delta_{nm} = E_n - E_m$. Equation (A3) then transforms into Equation (6):

$$J_\mu = \sum_{n>m}\Delta_{nm}\,(\Gamma^{-}_{nm,\mu}P_n - \Gamma^{+}_{nm,\mu}P_m).\text{(A4)}$$

Equation (A4) carries a clear physical interpretation: the term $\Gamma^{-}_{nm,\mu}$ corresponds to the process in which the system transitions from a higher energy level $|\psi_n\rangle$ to a lower energy level $|\psi_m\rangle$, releasing energy to the thermal bath. Conversely, the term $\Gamma^{+}_{nm,\mu}$ represents the reverse process, where the system absorbs energy from the thermal bath.

# Appendix B Analytical Solution for Heat Flow when $N_l = N_r = 1$

If each spin ensemble contains only one qubit, only two qubits couple within the system. The system Hamiltonian can be expressed as:

$$\hat{H} = \frac{\varepsilon_l}{2}\hat{\sigma}^l_z + \frac{\varepsilon_r}{2}\hat{\sigma}^r_z + \frac{\lambda}{4}\hat{\sigma}^l_x\hat{\sigma}^r_z,\text{(B1)}$$

Here, $\varepsilon_l$ and $\varepsilon_r$ denote the spin-splitting energies; $\hat{\sigma}_x, \hat{\sigma}_y, \hat{\sigma}_z$ represent the Pauli matrices; and $\lambda$ indicates the coupling strength. In the two-qubit system, we consider the four basis vectors $|0,0\rangle, |0,1\rangle, |1,0\rangle, |1,1\rangle$ to derive the eigenvalues and eigenstates of the Hamiltonian. By setting $\varepsilon_\mathrm{l} < \varepsilon_\mathrm{r}$, we obtain the eigenstates and their corresponding eigenvalues as follows:

$$|s_1\rangle = \cos\frac{\theta}{2}|1,1\rangle + \sin\frac{\theta}{2}|0,1\rangle, E_1 = \frac{\varepsilon_\mathrm{r} + \sqrt{\varepsilon_\mathrm{l}^2 + \lambda^2/4}}{2}, \quad \text{(B2a)}$$

$$|s_2\rangle = -\sin\frac{\theta}{2}|1,1\rangle + \cos\frac{\theta}{2}|0,1\rangle, E_2 = \frac{\varepsilon_\mathrm{r} - \sqrt{\varepsilon_\mathrm{l}^2 + \lambda^2/4}}{2}, \quad \text{(B2b)}$$

$$|s_3\rangle = \cos\frac{\theta}{2}|1,0\rangle - \sin\frac{\theta}{2}|0,0\rangle, E_3 = \frac{-\varepsilon_\mathrm{r} + \sqrt{\varepsilon_\mathrm{l}^2 + \lambda^2/4}}{2}, \quad \text{(B2c)}$$

$$|s_4\rangle = \sin\frac{\theta}{2}|1,0\rangle + \cos\frac{\theta}{2}|0,0\rangle, E_4 = \frac{-\varepsilon_\mathrm{r} - \sqrt{\varepsilon_\mathrm{l}^2 + \lambda^2/4}}{2}, \quad \text{(B2d)}$$

where $\tan\theta = \lambda/2\varepsilon_\mathrm{l}$. In the weak-coupling regime, $\lambda \ll \{\varepsilon_\mathrm{l}, \varepsilon_\mathrm{r}\}$, and hence $\theta \to 0$. Expanding $\cos\theta$ and $\sin\theta$ for small $\theta$, we obtain $\cos\theta \approx 1 - \theta^2/2, \sin\theta \approx \theta$. Based on Eq. (B2), the transition coefficients between energy levels are obtained as follows: $\langle s_1|\hat{J}_x^\mathrm{l}|s_2\rangle^2 = \langle s_3|\hat{J}_x^\mathrm{l}|s_4\rangle^2 = \frac{1}{4}(1-\theta^2)$, $\langle s_1|\hat{J}_x^\mathrm{r}|s_3\rangle^2 = \langle s_2|\hat{J}_x^\mathrm{r}|s_4\rangle^2 = \frac{1}{4}(1-\theta^2)$, and $\langle s_1|\hat{J}_x^\mathrm{r}|s_4\rangle^2 = \langle s_2|\hat{J}_x^\mathrm{r}|s_3\rangle^2 = \theta^2/4$. Similarly, the energy differences between the transitioning levels can be calculated as $\Delta_{12} = \Delta_{34} \approx \varepsilon_\mathrm{l}(1+\theta^2/2)$, $\Delta_{13} = \Delta_{24} \approx \varepsilon_\mathrm{r} \Delta_{14} \approx \varepsilon_\mathrm{r} + \varepsilon_\mathrm{l}(1+\theta^2/2)$, and $\Delta_{23} \approx \varepsilon_\mathrm{r} - \varepsilon_\mathrm{l}(1+\theta^2/2)$. Thus, the heat current $J_\mathrm{r}$ can be expressed as

$$\begin{aligned} J_\mathrm{r} = \quad & \Delta_{12}(\Gamma_{12}^- P_1 - \Gamma_{12}^+ P_2) + \Delta_{34}(\Gamma_{34}^- P_3 - \Gamma_{34}^+ P_4) \\ & + \Delta_{14}(\Gamma_{14}^- P_1 - \Gamma_{14}^+ P_4) + \Delta_{23}(\Gamma_{23}^- P_2 - \Gamma_{23}^+ P_3). \end{aligned} \quad \text{(B3)}$$

We expand the transition rates and occupation probabilities in $\theta^2$, i.e., $\Gamma = \Gamma^{(0)} + \Gamma^{(1)}$, and $P = P^{(0)} + P^{(1)}$, where the first-order corrections are of order $\theta^2$. Retaining terms up to first order in $\theta^2$, we obtain

$$\begin{aligned} J_\mathrm{r} \approx \quad & \Delta_{12}^{(0)}(\Gamma_{12}^{-(0)} P_1^{(1)} - \Gamma_{12}^{+(0)} P_2^{(1)} + \Gamma_{12}^{-(1)} P_1^{(0)} - \Gamma_{12}^{+(1)} P_2^{(0)}) \\ & + \Delta_{34}^{(0)}(\Gamma_{34}^{-(0)} P_3^{(1)} - \Gamma_{34}^{+(0)} P_4^{(1)} + \Gamma_{34}^{-(1)} P_3^{(0)} - \Gamma_{34}^{+(1)} P_4^{(0)}) \\ & + \Delta_{14}^{(0)}(\Gamma_{14}^- P_1^{(0)} - \Gamma_{14}^+ P_4^{(0)}) + \Delta_{23}^{(0)}(\Gamma_{23}^- P_2^{(0)} - \Gamma_{23}^+ P_3^{(0)}). \end{aligned} \quad \text{(B4)}$$

Combining Eq. (4) and Eq. (B4), we obtain the steady-state current expression shown in Eq. (7).

# Appendix C Analytical Solution for Heat Current in the

## Large-Spin Limit

For large spin numbers $N_\mu$ and in the low-excitation regime $(\langle \hat{a}_\mu^\dagger \hat{a}_\mu \rangle \ll N_\mu)$, the Holstein-Primakoff (HP) transformation allows us to approximate the mapping of spin operators to bosonic operators as $\hat{J}_+^\mu \approx \sqrt{N_\mu}\hat{a}_\mu^\dagger$, $\hat{J}_-^\mu \approx \hat{a}_\mu\sqrt{N_\mu}$, and $\hat{J}_z^\mu = \hat{a}_\mu^\dagger \hat{a}_\mu - N_\mu/2$. Here, the bosonic operators satisfy the commutation relation $[\hat{a}_\mu, \hat{a}_\nu^\dagger] = \delta_{\mu\nu}$. Using the spin operator relation $\hat{J}_x^\mu = (\hat{J}_+^\mu + \hat{J}_-^\mu)/2$ and substituting the aforementioned HP transformation into the Hamiltonian in Eq. (2), we obtain the following approximation

$$\hat{H}_\mathrm{S} \approx \varepsilon_i(\hat{a}_\mathrm{l}^\dagger \hat{a}_\mathrm{l} - \frac{N_\mathrm{l}}{2}) + \varepsilon_\mathrm{r}(\hat{a}_\mathrm{r}^\dagger \hat{a}_\mathrm{r} - \frac{N_\mathrm{r}}{2}) + \frac{\lambda\sqrt{N_\mathrm{l}}}{2}(\hat{a}_\mathrm{l}^\dagger + \hat{a}_\mathrm{l})(\hat{n}_\mathrm{r} - \frac{N_\mathrm{r}}{2}). \text{(C1)}$$

To eliminate the linear interaction term involving $\hat{a}_\mathrm{l}$ in $\hat{H}_\mathrm{S}$, we introduce the displacement operator $\hat{D}(\alpha) = \exp[\alpha(\hat{a}_\mathrm{l}^\dagger - \hat{a}_\mathrm{l})]$. This operator acts on the system as $\tilde{\hat{H}}_\mathrm{S} = \hat{D}^\dagger(\alpha)\hat{H}_\mathrm{S}\hat{D}(\alpha)$, where the displacement parameter is $\alpha = -\lambda\sqrt{N_\mathrm{l}}(m_\mathrm{r} - N_\mathrm{r}/2)/2\varepsilon_\mathrm{l}$. The effective Hamiltonian $\tilde{\hat{H}}_\mathrm{S}$ after the displacement transformation becomes

$$\tilde{\hat{H}}_\mathrm{S} = \varepsilon_\mathrm{l}\hat{a}_i^\dagger \hat{a}_i + \varepsilon_\mathrm{r}\hat{a}_\mathrm{r}^\dagger \hat{a}_\mathrm{r} - \frac{\lambda^2 N_\mathrm{l}(\hat{n}_\mathrm{r} - \frac{N_\mathrm{r}}{2})^2}{4\varepsilon_\mathrm{l}}. \text{(C2)}$$

Correspondingly, the eigenstates of the system can be described as displaced Fock states $|\psi_{m_\mathrm{l},m_\mathrm{r}}\rangle = |\phi_{m_\mathrm{l}}\rangle_{m_\mathrm{r}} \otimes |m_\mathrm{r}\rangle$, where $|\phi_{m_\mathrm{l}}\rangle_{m_\mathrm{r}} = D(\alpha)|m_\mathrm{l}\rangle$. The eigenvalues of the system are given by $E_{m_\mathrm{l},m_\mathrm{r}} = \varepsilon_\mathrm{l} m_\mathrm{l} + \varepsilon_\mathrm{r} m_\mathrm{r} - \frac{\lambda^2 N_\mathrm{l}(m_\mathrm{r} - N_\mathrm{r}/2)^2}{4\varepsilon_\mathrm{l}}$. In the weak-coupling $(\lambda/\varepsilon_\mu \ll 1)$ and low-excitation regime, the interaction term simplifies to $\lambda^2 N_\mathrm{l}(m_\mathrm{r} - N_\mathrm{r}/2)^2/2\varepsilon_\mathrm{l} \approx \lambda^2 N_\mathrm{l} N_\mathrm{r}^2/8\varepsilon_\mathrm{l}$. Considering the incoherent transition rates in Eq. (5) and applying the Holstein-Primakoff (HP) transformation $\hat{J}_x^\mu = (\hat{J}_+^\mu + \hat{J}_-^\mu)/2 = \sqrt{N_\mu}(\hat{a}_\mu + \hat{a}_\mu^\dagger)/2$, we obtain

$$\Gamma_\mathrm{l}^\pm(\Delta_{m_\mathrm{l}-1,m_\mathrm{r}-1}^{m_\mathrm{l},m_\mathrm{r}}) = \frac{\gamma_\mathrm{l}(\pm\varepsilon_\mathrm{l})m_\mathrm{l}(\pm\varepsilon_\mathrm{l})N_\mathrm{l}m_\mathrm{l}}{4}, \text{(C3a)}$$

$$\Gamma_\mathrm{r}^\pm(\Delta_{m_{\mathrm{l}'},m_\mathrm{r}-1}^{m_\mathrm{l},m_\mathrm{r}}) = \frac{N_\mathrm{r}m_\mathrm{r}}{4}\gamma_\mathrm{r}[\pm(\varepsilon_\mathrm{l}m_\mathrm{l} - \varepsilon_\mathrm{l}m_{\mathrm{l}'} + \varepsilon_\mathrm{r})]m_\mathrm{r} \times [\pm(\varepsilon_\mathrm{l}m_\mathrm{l} - \varepsilon_\mathrm{l}m_{\mathrm{l}'} + \varepsilon_\mathrm{r})]|\ {}_{m_\mathrm{r}}\langle\psi_{m_\mathrm{l}}|\psi_{m_{\mathrm{l}'}}\rangle_{m_\mathrm{r}-1}|^2. \text{(C3b)}$$

Let the coupling coefficient be defined as $D_{m_l,m_{l'}}(\alpha) = {}_{m_r}\langle\psi_{m_l}|\hat{D}(\alpha)|\psi_{m_{l'}}\rangle_{m_r-1}$. In the weak coupling regime where $\frac{\lambda\sqrt{N_l}}{2\varepsilon_l} \ll 1$, we expand the displacement operator to obtain the approximation

$$\hat{D}_{m_l,m_{l'}}\left(\frac{\lambda\sqrt{N_l}}{2\varepsilon_l}\right) \approx \delta_{m_l,m_{l'}} + \frac{\lambda\sqrt{N_l}}{2\varepsilon_l}\sqrt{m_{l'}+1}\,\delta_{m_l,m_{l'}+1} - \frac{\lambda\sqrt{N_l}}{2\varepsilon_l}\sqrt{m_{l'}}\,\delta_{m_l,m_{l'}-1}. \tag{C4}$$

Transitions are considered when $\varepsilon_l \leqslant \varepsilon_r$. The transition rate comprises quasi-elastic, sum-frequency, and difference-frequency terms. This is expressed as

$$\begin{aligned}\Gamma_r^{\pm}(\Delta^{m_l,m_r}_{m_{l'},m_r-1}) &= \gamma_r(\pm\varepsilon_r)n_r(\pm\varepsilon_r)\frac{N_r m_r}{4} \\ &+\gamma_r(\pm(\varepsilon_r+\varepsilon_l))n_r(\pm(\varepsilon_r+\varepsilon_l))\frac{N_r m_r}{4}\frac{\lambda^2 N_l(m_l+1)}{4\varepsilon_l^2}\delta_{m_l,m_{l'}+1} \\ &+\gamma_r(\pm(\varepsilon_r-\varepsilon_l))n_r(\pm(\varepsilon_r-\varepsilon_l))\frac{N_r m_r}{4}\frac{\lambda^2 N_l m_l}{4\varepsilon_l^2}\delta_{m_l,m_{l'}-1}.\end{aligned} \tag{C5}$$

We expand the transition rates and occupation probabilities perturbatively in $\left(\frac{\lambda\sqrt{N_l}}{2\varepsilon_l}\right)^2$, namely

$$\Gamma \approx \Gamma^{(0)} + \left(\frac{\lambda\sqrt{N_l}}{2\varepsilon_l}\right)^2\Gamma^{(1)}, P \approx P^{(0)} + \left(\frac{\lambda\sqrt{N_l}}{2\varepsilon_l}\right)^2 P^{(1)},$$

Combining this with the dressed-state master equation in Eq. (4), the steady-state current can be expressed as

$$\begin{aligned}J_r \approx \left(\frac{\lambda\sqrt{N_l}}{2\varepsilon_l}\right)^2 \times \Big\{&\varepsilon_r \sum_{m_l,m_r}\big[\Gamma_r^{-}(\Delta^{m_l+1,m_r+1}_{m_l,m_r})P^{(0)}_{m_l+1,m_r+1} \\ &-\Gamma_r^{+}(\Delta^{m_l+1,m_r+1}_{m_l,m_r})P^{(0)}_{m_l,m_r}\big] \\ &-\varepsilon_l \sum_{m_l,m_r}\big[\Gamma_r^{-}(\Delta^{m_l-1,m_r+1}_{m_l,m_r})P^{(0)}_{m_l-1,m_r+1} \\ &-\Gamma_r^{+}(\Delta^{m_l-1,m_r+1}_{m_l,m_r})P^{(0)}_{m_l,m_r}\big]\Big\},\end{aligned} \tag{C6}$$

Here, the zeroth-order occupation probability is given by $P^{(0)}_{m_l,m_r} = P_{m_l} \times P_{m_r}, P_{m_\mu} = e^{-\beta_\mu\varepsilon_\mu m_\mu}/Z_\mu$, with the partition function defined as $Z_\mu = \sum_{m_\mu=0}^{\infty} e^{-\beta_\mu\varepsilon_\mu m_\mu}$, where $m_\mu$ denotes the boson excitation number. Furthermore, using Eq. (C5), we obtain $\Gamma_r^{\pm}(\Delta^{m_l+1,m_r+1}_{m_l,m_r}) = \gamma_r(\pm(\varepsilon_r+\varepsilon_l))n_r(\pm(\varepsilon_r+\varepsilon_l))\frac{N_r}{4}(m_r+1)(m_l+1)$ and $\Gamma_r^{\pm}(\Delta^{m_l-1,m_r+1}_{m_l,m_r}) = \gamma_r(\pm(\varepsilon_r-\varepsilon_l))n_r(\pm(\varepsilon_r-\varepsilon_l))\frac{N_r m_l}{4}(m_r+1)$, yielding the steady-state current in Eq. (8).